\documentclass[aps,prl,showpacs,address,twocolumn,letter]{revtex4-1}
\usepackage{amsfonts}
\usepackage{amsmath}
\usepackage{amssymb}
\usepackage{graphicx}
\usepackage{subfigure}
\usepackage{txfonts}
\begin{document}

\title{Quantum anomalous Hall effect and tunable topological states in $3d$
transition metals doped silicene}
\author{Xiao-Long Zhang}
\affiliation{Beijing National Laboratory for Condensed Matter Physics, Institute of
Physics, Chinese Academy of Sciences, Beijing 100190, China}
\author{Lan-Feng Liu}
\affiliation{Beijing National Laboratory for Condensed Matter Physics, Institute of
Physics, Chinese Academy of Sciences, Beijing 100190, China}
\author{Wu-Ming Liu}
\affiliation{Beijing National Laboratory for Condensed Matter Physics, Institute of
Physics, Chinese Academy of Sciences, Beijing 100190, China}
\date{\today }
\begin{abstract}
We engineer quantum anomalous Hall effect in silicene via doping $3d$ transition metals. We show that
there exists a stable quantum anomalous Hall effect in Vanadium doped
silicene using both analytical model and Wannier interpolation. We also
predict the quantum valley Hall effect and electrically
tunable topological states could be realized in certain transition metal doped silicene where the energy band
inversion occurs. Our finding provides new scheme for the realization of
quantum anomalous Hall effect and platform for electrically controllable topological states.
\end{abstract}

\maketitle

$Introduction.-$The recently discovered topological insulators (TIs) \cite%
{TIRMP,TIRMP2,KM1,KM2,BHZ} have aroused great interest in the fields of
condensed matter physics and materials science due to the
metallic boundary states protected by time-reversal symmetry (TRS). TIs have
also become perfect breeding ground for a variety of exotic quantum
phenomena \cite{TIRMP,TIRMP2}. In particular, breaking the TRS respected by TIs via
magnetic doping is predicted to give birth to the Majorana fermion \cite{Majorana},
topological magnetoelectric effect \cite{topoele}, and the so-called quantum
anomalous Hall effect (QAHE) \cite{Qi.X.L, Liu.C.X,Yu,Ding,Hongbin}, which has quantised Hall
conductance in the absence of external field \cite{Haldane} and can be
intuitively thought as half of TIs. Since TIs have been fabricated in materials ranging from
2D \cite{HgTe} to 3D \cite{Bi2Se3}, engineering these novel
phenomena in real materials represents one of the most fascinating areas in this field.

As the 2D TI (also known as quantum spin Hall (QSH) insulator) graphene \cite{graphene,KM1,KM2} has been
altering the research direction of nanoelectronics from silicon-based
materials to carbon-based ones, however, the advent of silicene \cite%
{silicene,exp1,exp2,exp3,exp4}, which is the silicon equivalent for
graphene, seems to turn the tide. Silicene is closely analogous to graphene
in the sense that it consists of a single layer of Si atoms arranged to a
low buckled honeycomb lattice, and its low energy physics can be described
by Dirac-type energy-momentum dispersion akin to that in graphene \cite{silicene}, hence the inherited many
intriguing  properties, including the expected Dirac fermions and QSH effect \cite{LCCL}. Yet a striking difference between silicene and
graphene is that the stable silicene monolayer has additional buckling degree
\cite{silicene}, which accounts for the relatively large (1.55meV) spin orbit coupling (SOC)
induced gap \cite{LCCL} in silicene and a couple of unusual quantum
phenomena recently reported \cite{Ezawa,kaust,super}. Indeed, these features together with the natural compatibility with current silicon-based microelectronics industry make silicene a promising candidate for future nanoelectronics application. Moreover, from the view of practical applications, it is highly appreciated if magnetism or sizable band gap or both, like in QAHE with additional edge states protected by topology, can be established in the nonmagnetic silicene, especially in the presence of the bulking degree.

In this Letter, we present a systematic investigation of the adsorption properties and
magnetism of $3d$\ transition metals (TM) doped silicene. We demonstrate
that $3d$\ TM strongly bonding with silicene and the TM-silicene systems
are strongly magnetic. From combined tight-binging (TB) model analysis
and first-principles Wannier interpolation, we investigate the topological
properties of the resulting TM-silicene systems. Our results suggest the V
doped silicene hosts a stable QAHE, and this system can also be half-metallic \cite{halfmetal} if the Fermi level is properly tuned. Further, a close study of the TB model in the
band inverted regime gives rise to another topologically nontrivial
state, which supports quantum valley Hall effect (QVHE) \cite{QVHE}. We predict the
resulting QAHE and QVHE can be tuned directly using an external electrical
field.

$Adsorption$ $and$ $magnetism$ $analysis.-$ We use $%
4\times 4$ supercell of silicene to model the interaction between $3d$ TM
(Sc, Ti, V, Cr, Mn, Fe, Co, Ni) and silicene. As silicene has
buckled geometry, we consider three high
symmetry adsorption sites, namely the hollow (H) site at the center of a hexagon, two
top sites denoted as T$_{A}$ and T$_{B}$ corresponding to the top of Si
atoms belonging to A and B sublattice, respectively (see Fig. 1). To evaluate
the effect of on-site Coulomb interactions among $3d$ electrons of adatoms
on the equilibrium structure and magnetic properties of the TM-silicene
system, the simulations have been carried out within generalized gradient
approximation (GGA) \cite {PBE} and GGA$+U$ \cite{Uvalue} framework separately.

\begin{figure}[tbph]
\includegraphics[width = 6.5cm]{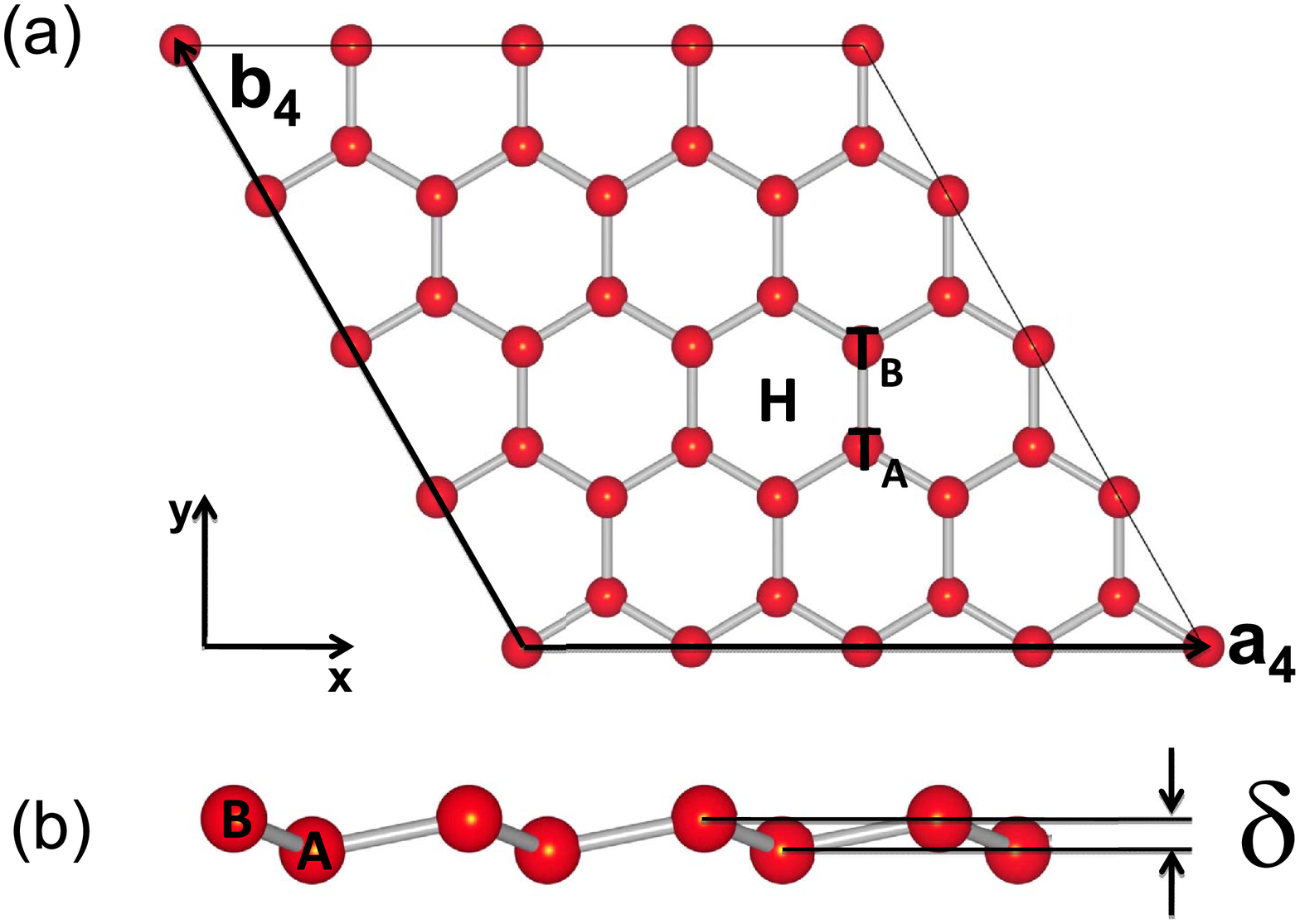}
\caption{{}The lattice geometry of $3d$ transition metals doped silicene
with lattice constant $|a_{4}|=|b_{4}|=4a$, where $a=3.86$ $\mathring{A}$ is
the lattice constant of silicene. (a) Top view of $4\times 4$ silicene
monolayer where the 3 adsorption sites (Hollow (H), top of A sublattice (T$_{A})$ and B sublattice (T$_{B}))$ are marked out with black letters. (b)
Side view of silicene, the two equivalent Si sublattices are labeled as A
and B, respectively, with a buckled distance $\protect\delta =0.44$ $\mathring{A}.$}
\end{figure}

Let us first focus on the GGA case. From our first-principles results, of
the three adsorption sites concerned, all $3d$ TM energetically favor H
site, which is $0.02$ eV $\sim 0.60$ eV and $0.20$ eV$\sim 0.80$ eV higher
in adsorption energy ($\Delta E$) \cite{deltaE} than T$_{A}$ and T$_{B}$ site, respectively (see Fig. 2 in
Ref. \cite{supp}). The bondings between $3d$ TM and silicene are strongly covalent as
manifested by a much larger $\Delta E$ ranging from 2.44 eV to 4.75
eV. The unusual large $\Delta E$ compared with that in graphene case \cite%
{graphenedelta, Ding} could be related to the covalently more active $sp^{3}$-like orbitals of
silicene, which result from the unique buckled geometry .

Much like the graphene case \cite{graphenedelta, Ding}, most of TM (except Ni) doped silicene
exhibit magnetism with sizable magnetic moments ranging from $\sim 1$ $\mu
_{B}$ to $\sim 5$ $\mu _{B}$. A relatively large magnetic moment is key to
the realization of QAHE in silicene which we will discuss later. We also
note that when some TM (Sc, Ti, Cr) adsorbing on H site, the density of
states (DOS) show peaks at the Fermi level, indicating that these systems
could be magnetic instable and may undergo Jahn-Teller distortion
to lower total energy (especially for Sc, see Ref. \cite{supp} for
detailed discussion).

The resulting magnetic moments and possible Jahn-Teller distortion
aforementioned can be understood in the light of symmetry considerations.
When TM are deposited on high symmetry sites of \!silicene (H, T$_{A}$, T$_{B}$%
), the $3d$ \!subshell of adatom split into three groups under the $C_{3v}$
symmetric crystal field of system: the $3d_{3z^{2}-r^{2}}$ state
corresponding to $A_{1}$ symmetry group, the twofold degenerate $E_{1}$
group consisting of the $3d_{xz}$ and $3d_{yz}$ states and the $E_{2}$
\!group consisting of the $3d_{xy}$ and $3d_{x^{2}-y^{2}}$ states.
Therefore, \!the three groups of $3d$ states hybridise with $\pi$
orbitals of silicene weakly or strongly according to the different
symmetrical properties in similar way as in Benzene \cite{hmweng} and
graphene cases \cite{Wehling}. Since 3d orbitals are anti-bonding states, the
energy order of them are usually $\epsilon(E_{2})\!<\!\epsilon
(A_{1})\!<\!\epsilon(E_{1})$ and in general $\epsilon \left( E_{2}\right) $ and
$\epsilon (A_{1})$ are close to each other due to similar hybridization
strength with $\pi$ orbitals. After incorporating spin
polarization, the 3 groups of states split, according to different splitting
energy, into 10 spin polarized orbitals. Meanwhile, the outer $4s$
electrons of adatoms experience relatively large electrostatic interaction
from the $\pi$ manifold of silicene than $3d$ shells due to its spherical
symmetry and delocalized nature, making possible charge transfer from the $%
4s$ to $3d$ shells.\! Thus, the crystal field splitting, spin
splitting, together with the occupation number, mostly dominate the
electronic structure of adsorbed TM ions (see Ref. \cite{supp} for detailed discussion).\!

Since the strong correlation effect of $3d$ electrons is not negligible for
a practical description of adsorption, we next consider GGA$+U$ case. After full relaxation, the adsorption geometry of adatom-silicene system
is strongly altered compared with GGA case (see Fig. 3(b) in Ref. \cite{supp}). And most $3d$ TM still favor H site (except Mn, which now
energetically favors T$_{A}$ site). Clearly, the geometry change of adatom-silicene systems are the
direct consequence of on-site Coulomb interactions among $3d$ electrons, which modify the
electron distribution in $3d$ and $4s$ shells of adatoms and $\pi $ orbitals
of silicene as can be seen from the changes of magnetic moments (see Ref. \cite{supp} for detailed discussion).

$Effective$ $model$ $and$ $Chern$ $number$ $analysis.-$In this part, we turn
to the main finding of this work, namely the QAHE in the absence of external
field via doping $3d$ TM and the prediction of electrically tunable
topological states. As has been shown in Refs. \cite{Ding,Hongbin}, the QAHE could be realized via doping certain $3d$ or $5d$
adatoms on the hollow site of graphene. In Fe doped graphene case \cite{Ding}, the QAHE gaps occurs around the Dirac $%
K$ points of the Brillouin zone, and the low energy physics can be described
by a Hamiltonian for graphene in the presence of extrinsic Rashba SOC ($%
\lambda _{R}^{ext}$) and exchange field ($M$) \cite{Qiao rapid} introduced
solely by the adatom. Here in silicene, however, owing to the low buckled
structure, there exists intrinsic Rashba SOC ($\lambda _{R}^{int}$) \cite%
{LCCB}. Moreover, when depositing $3d$\ TM on the stable adsorption site,
the induced inequality of AB sublattice potential ($\Delta$) necessarily
arise and compete with magnetization. Below we identify
conditions for the realization of QAHE in silicene based on a
effective Hamiltonian \cite{LCCB} by introducing a staggered AB sublattice
potential besides SOC ($\lambda _{R}^{ext}$ and $\lambda _{so}$) and
exchange field ($M$).
\begin{figure}[tbph]
\centering{\subfigure{
     \includegraphics[width=8cm]{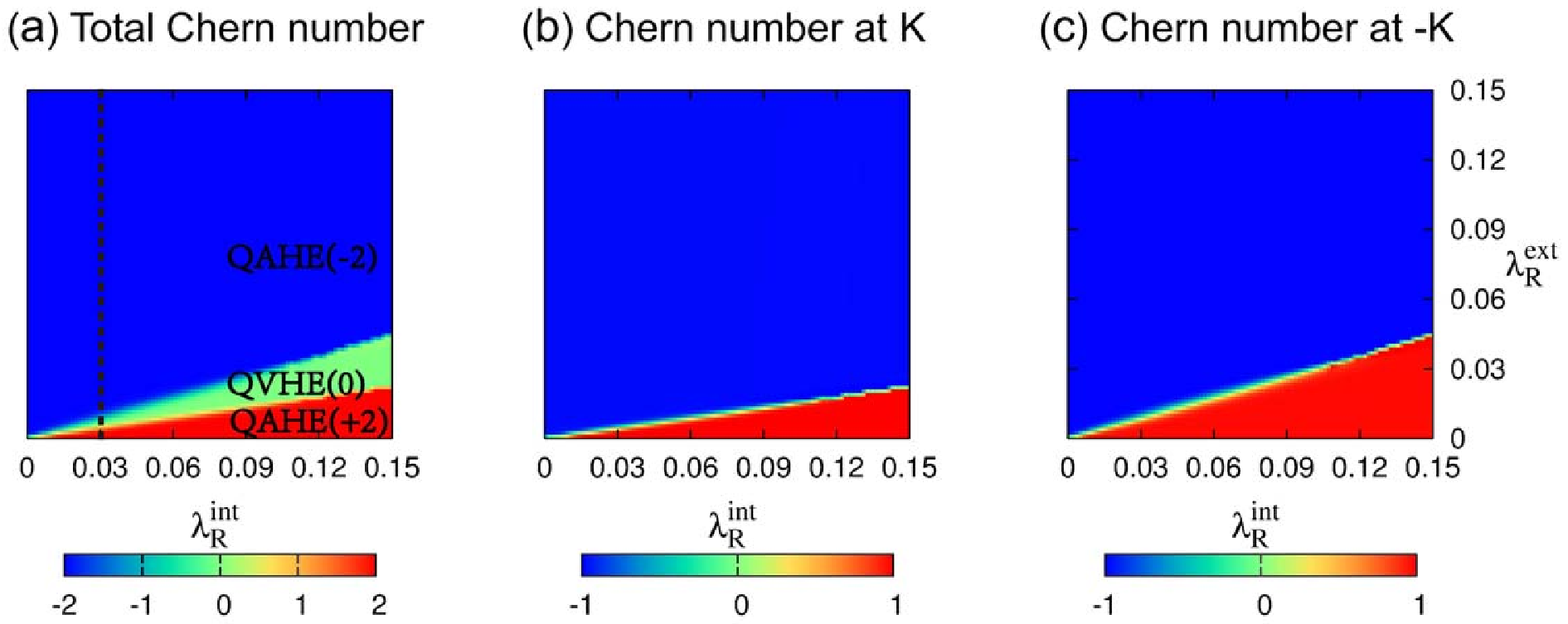}} } \centering{%
\subfigure{
     \includegraphics[width=7.5cm]{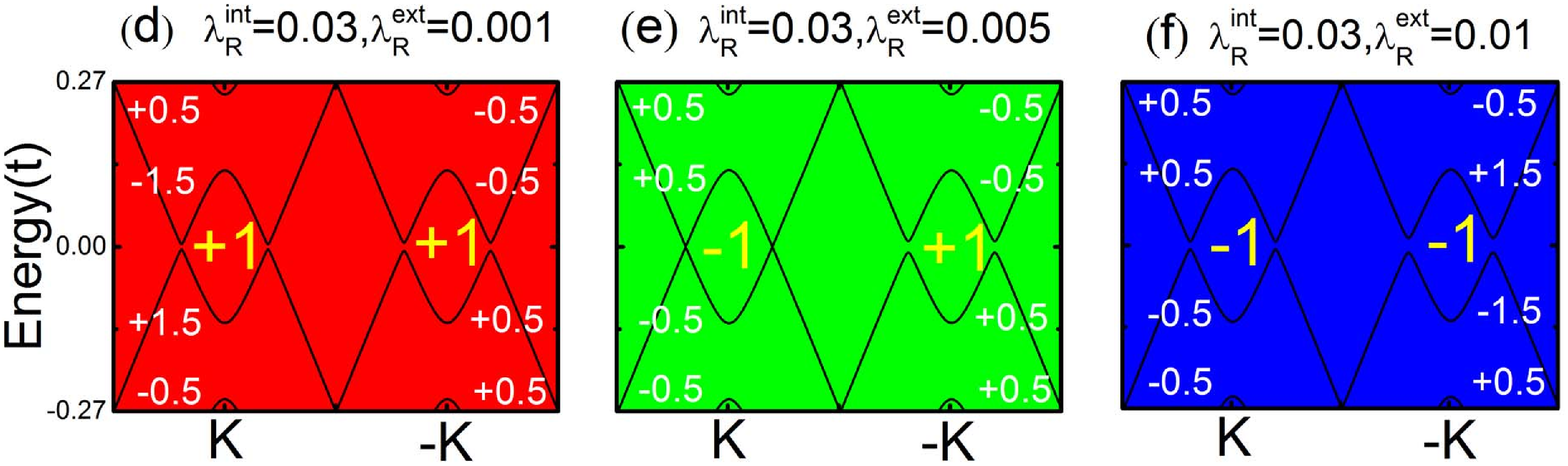}} } \centering{\
\subfigure{
     \includegraphics[width=7.2cm]{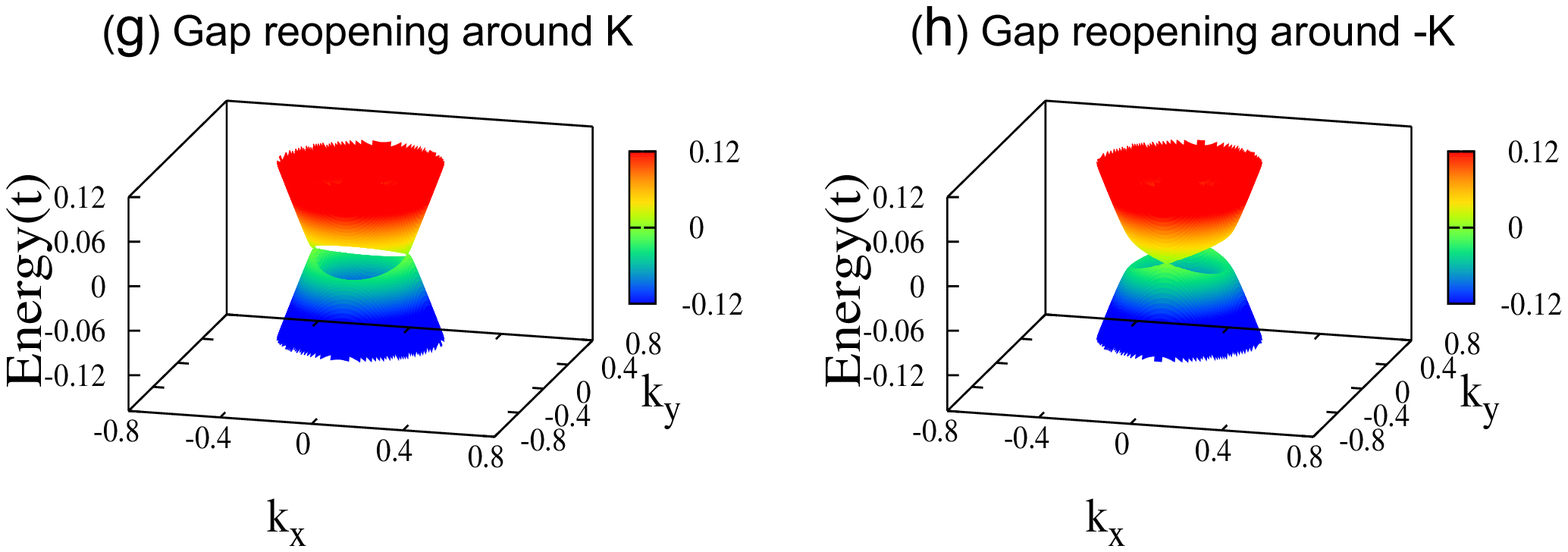}} }
\caption{The transition of Chern number by tuning $\protect\lambda _{R}^{ext}$ and $\protect\lambda _{R}^{int}$ (in unit of $t$). (a) Three topological
nontrivial states, QAHE(2), QVHE(0) and QAHE(-2) with Chern number $+2$, $0$, and $-2$, can be obtained from different combination of $\protect\lambda _{R}^{ext}$ and $\protect\lambda _{R}^{int}$. (b) and (c) represent the
the variation of $\mathcal C_{K}$ and $\mathcal C_{-K}$. (d), (e) and (f) depict
the band structure along $k_{y}=0$ line in $BZ$ for the three topological
states (QHE(2), QVHE(0) and QHE(-2)) in (a). The yellow integers ($\pm1$) represent $\mathcal C_{K}$ and $\mathcal C_{-K}$, corresponding to the sum of topological charge of each valence bands (the white $\pm0.5$ and $\pm1.5$). (g) and (h) show
the gap closing around $K$ and $-K$. They are the transition states
from QAHE(+2) to QVHE(0) and from QVHE(0) to QAHE(-2), respectively.}
\end{figure}
In the basis of $\{A,B\}\otimes \{\uparrow ,\downarrow \}$, the Hamiltonian
reads:%
\begin{equation}
H_{eff}^{\pm }=H_{s}^{\pm }+H_{d}^{\pm },
\end{equation}%
with
$H_{s}^{\pm }=\varepsilon _{eff}\tau _{0}\otimes \sigma _{0}\pm \tau
_{3}\otimes h_{11}+\hbar v_{F}(k_{x}\tau _{1}\mp k_{y}\tau _{2})\otimes
\sigma _{0}$,
$H_{d}^{\pm }=\lambda _{R}^{ext}(\pm \tau _{1}\otimes \sigma _{2}-\tau
_{2}\otimes \sigma _{1})+\Delta \tau _{3}\otimes \sigma _{0}+M\tau
_{0}\otimes \sigma _{3}$,
$h_{11}=-\lambda _{so}\sigma _{3}-a\lambda _{R}^{int}(k_{y}\sigma
_{1}-k_{x}\sigma _{2})$,
where $H_{eff}^{\pm }$ are the total Hamiltonian for the two inequivalent
Dirac points $K\left( +\right) $ and $-K\left( -\right) $, $H_{s}^{\pm }$
are the low energy effective Hamiltonian for the QSH insulator silicene, $%
H_{d}^{\pm }$ include all the effects introduced by the $3d$ dopants,
including effective spin-dependent magnetic field $M$, site-dependent
staggered potential $\Delta $ and the resulting extrinsic Rashba SOC $%
\lambda _{R}^{ext}$. The $\mathbf{\tau }$ and $\mathbf{\sigma }$ are the
Pauli matrices acting separately on pseudospin (sublattice) and spin space, $%
\varepsilon _{eff}$ stands for the $\varepsilon _{1}-\lambda _{2nd}$ term in
Ref. \cite{LCCL}, $v_{F}$ and $a$ are the Fermi velocity and the lattice constant,
respectively, and $\lambda _{so}$ is the effective SOC.

\begin{figure*}[tbph]
\centering{\subfigure[]{
\includegraphics[width = 9.5cm,height = 5cm]{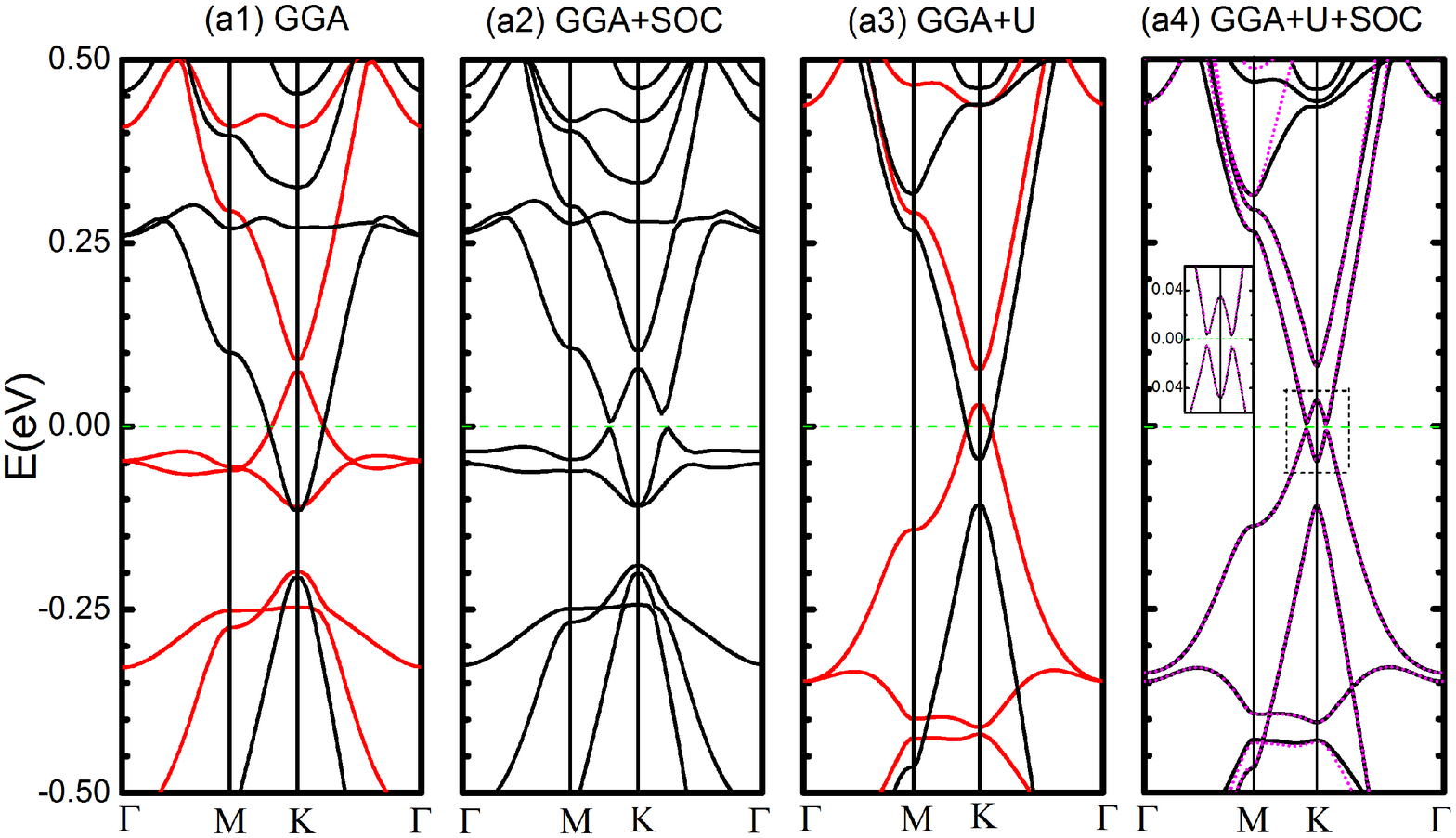}}}
\centering{\subfigure[]{
\includegraphics[width = 6.5cm,height = 4.7cm]{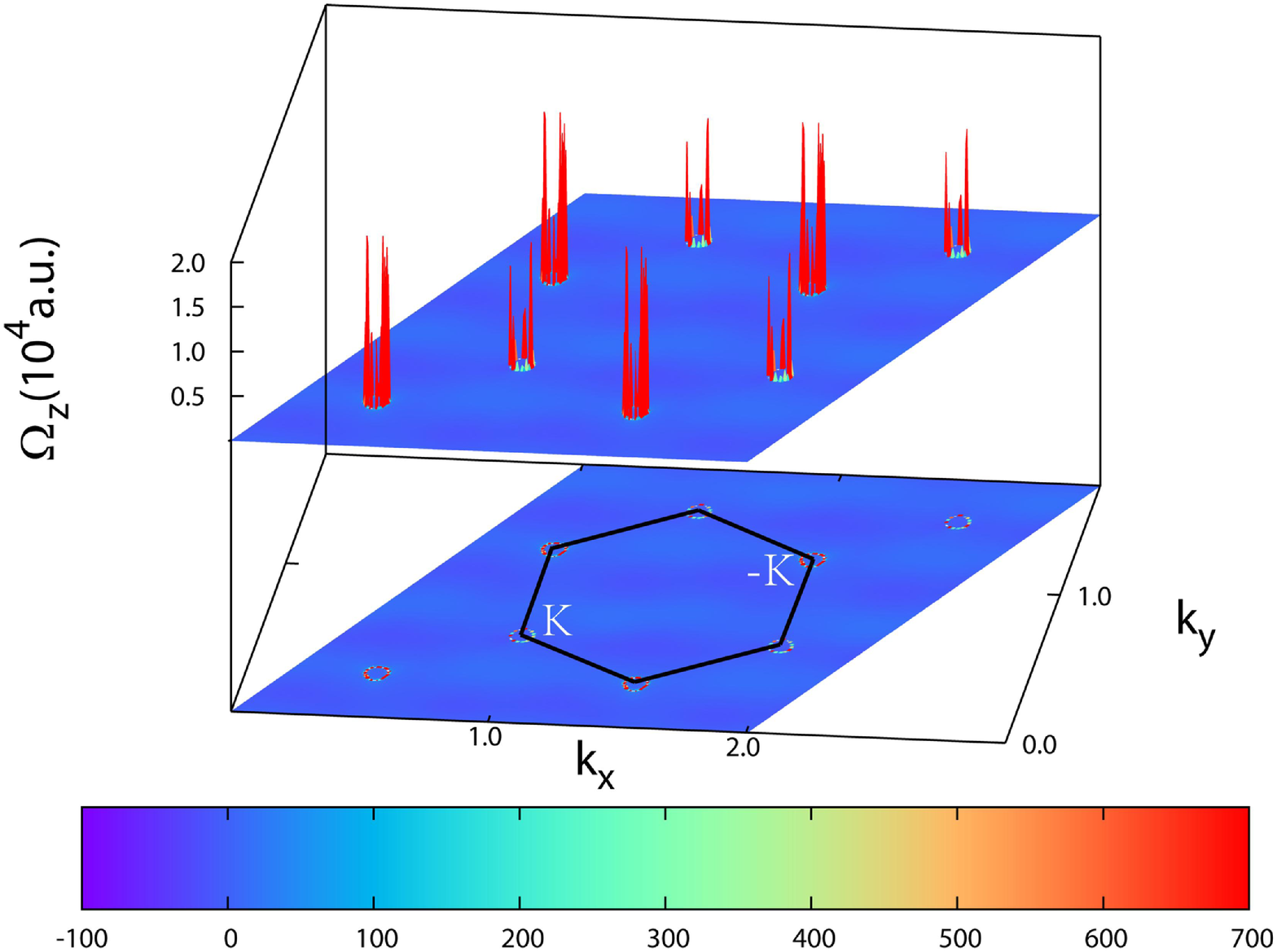}}}
\caption{{}(a) The band structures of V doped silicene from GGA ((a1)-(a2)) and GGA$%
+U$ ((a3)-(a4)), respectively. The red (black) color in (a1) and (a3)
correspond to majority spin (minority spin) subbands. After including SOC
effect, a gap is opened at the Fermi level ((a2) and (a4)). In (a4), the band
structure from Wannier interpolation is also shown in pink dashed lines. (b) The distribution of Berry curvature ($\Omega _{z}(k)$) in V doped silicene from GGA+$U$+SOC. The $BZ$ is marked
out with black hexagon.}
\end{figure*}
By diagonalizing the above Hamiltonian, the electronic structure around each
valley ($K$ and $-K$) in Brillouin zone can be
obtained (the $\varepsilon _{eff}$ term can be safely ignored). From the interplay between exchange field ($M$) and
staggered potential ($\Delta$) (see Fig. 4 in Ref. \cite{supp}), we conclude that energy bands with opposite
spin intersect, which is essential to the QAHE in Fe doped graphene, only
when $M/\Delta >1$. This is different from that in Ref. \cite{Qiao rapid},
where $\Delta =0$ and there always exists 2 degenerate points around Dirac
point as long as $M\neq 0$.

We found that either the extrinsic or the intrinsic Rashba SOC would lead to
insulating state when starting from the case $M/\Delta >1$. To identify the
topological properties of the resulting insulating state, we resort to the
Chern number ($\mathcal C$) analysis \cite{TKNN}. The $\mathcal C$ can be obtained by the
integral over the first Brillouin zone ($BZ$): $\mathcal C=\frac{1}{2\pi }%
\int_{BZ}\Omega (\mathbf{k})d^{2}k$. The $\Omega (\mathbf{k})$ is the usual
Berry curvature of all occupied states \cite{BC}:%
\begin{equation}
\Omega _{z}(\mathbf{k})=-2\sum\limits_{n}\sum\limits_{m\neq n}f_{n}\text{Im}%
\frac{\langle \psi _{n\mathbf{k}}|v_{x}|\psi _{m\mathbf{k}}\rangle \langle
\psi _{m\mathbf{k}}|v_{y}|\psi _{n\mathbf{k}}\rangle }{(\epsilon _{m\mathbf{k%
}}-\epsilon _{n\mathbf{k}})^{2}},
\end{equation}%
where $f_{n}$ is the Fermi-Dirac distribution function for band $n$, $\psi
_{n\mathbf{k}}$ is the Bloch function of eigenenergy $\epsilon _{n\mathbf{k}%
} $, $v_{x},v_{y}$ are the velocity operators. And anomalous Hall
conductivity is readily given by $\sigma _{xy}=(e^{2}/h)\mathcal C$.
Interestingly, the extrinsic Rashba SOC ($\lambda _{R}^{ext}$) gives insulating state with $\mathcal C=+2$ while the intrinsic one ($\lambda _{R}^{int}$) leads to that with $\mathcal C=-2 $. One may expect that different Chern number state can be realized if
tuning the two types of Rashba SOC properly in experiments. Fig. $2(a)$
indicates\ that this is indeed the case, where $\mathcal C$ can take integer
value of $+2,0,-2$ with different combination of $\lambda _{R}^{int}$ and $%
\lambda _{R}^{ext}$. Careful study shows that the tunable $\mathcal C$
originates from different response of bulk gap to the 2 types of Rashba SOC (%
$\lambda _{R}^{int}$ and $\lambda _{R}^{ext}$) around $K$ and $-K$. When
increasing $\lambda _{R}^{ext} $ while keeping $\lambda _{R}^{int}$ fixed,
for example, at $0.03t$ $(t=1.6$ eV is the nearest neighbor hopping parameter
\cite{LCCB}) as shown in Fig. $2(a)$, we can see clearly the transition of
Chern number of each valley ($\mathcal C_{K}$ and $\mathcal C_{-K}$) from $+1$
to $-1$ but with different rate, i.e., $\mathcal C_{K}$ experiences a topological
transition earlier than $\mathcal C_{-K}$ (see Figs. 2(b)-2(c)). The
step change of $\mathcal C_{K}$ and $\mathcal C_{-K}$ is justified by the observation of
bands touching and gap reopening around each valley (Figs. 2(d)-2(f)).
Notice that the rotational symmetry of the effective Hamiltonians along $z$
direction in any angle is broken after bringing in the Rashba SOC terms, and hence the
band touching happens only on $k_y=0$ line in $BZ$ for valley $K$ (Fig. 2(g)) while $k_x=0$ line for valley $-K$ (Fig. 2(h)). Consequently, the system can be in QAHE phase (with $\mathcal C$ being $+2$ or $-2$)
or QVHE (with $\mathcal C$ being 0 and $\mathcal C_{K}=-\mathcal C_{-K}=-1$) depending on different value of $\lambda
_{R}^{ext},$ which is controllable through an external gate voltage.

The effective SOC $\left( \lambda _{so}\right) $, however, further breaks
the particle-hole symmetry of the above Hamiltonian, making energy bands
shift up $($at valley $K)$ or down $($at valley $-K)$ relative to the Fermi
level while leaving the topological charge of each valley unchanged. Hence,
as long as the shifting is small, the system is still insulating and the
above discussion of topological transition remain valid.

Some of the topologically non-trivial phases can be realized in $3d$ TM
doped silicene as predicted by our first-principles calculations. For
example, we notice that the opposite spin subbands cross around the
Fermi level in V, Cr, Mn doped silicene due to relatively large
magnetization. Meanwhile, for these systems the spin-up and spin-down subbands near Fermi level are also gapped (Fig. 3(a1) and Fig. 3(a3)), which makes these systems candidates for half-metallic materials if tuning Fermi level properly. We take V-silicene as a prototype and discuss in detail, as this system is insulating with an energy gap around $6m$eV (Fig. 3(a2)) when only the SOC is turned on.
We find the SOC induced band gap in V-silicene is stable even strong correlation effect
of V is considered by including effective $U$ value ranging from $0$ to $6$
eV (Fig. 3(a4)).

To obtain an accurate Berry curvature distribution and hence Chern number in
first-principles level in V doped silicene, the Wannier interpolation technique
have been used \cite{xinjie, wannier90}. The band structure from first
principle is well reproduced by interpolation as indicated in Fig. 3(a4).
In Fig. 3(b), the Berry curvature ($\Omega _{z}(k)$) in $k$-space is
explicitly shown. As we may observe, the most nonzero values of Berry
curvature distribute around the Dirac $K$ points by forming small circles,
where exactly the avoided crossing happens (Fig. 3(a4)). By the integral over
BZ, we indeed find the Chern number of all occupied bands equals to an
integer value of $+2,$ which signals the V doped silicene is in QAHE phase.

For the proposed QVHE state in former model analysis, we note that
the $\lambda _{R}^{int}$ is rather small (being about $4.4\times 10^{-4}t$)
\cite{LCCB}. Therefore, the QVHE region in Fig. $2(a)$ would be quite small
and it may not be easy to directly observe the QVHE from experiment. To tide
it over, putting V-silicene system or its counterpart of TM doped Germanium
on substrate may be a feasible solution.

$Conclusions.-$To summarize, we have demonstrated that the $3d$ TM doped silicene can be intriguing materials as manifested by the induced strong magnetic moments, potential half-metallic property as well as sizable topologically non-trivial gaps. These features have also been confirmed in the presence of strong correlation effect of $3d$ TM. Moreover, we predicted the emerging of electrically controllable topological states (QAHE and QVHE phases as
characterized by different Chern number) in certain TM-silicene systems
where the energy bands are being inverted. Our work may provide new
candidate for the long-sought QAHE and platforms to manipulate topological phase transition electrically.

X. L. Zhang acknowledges very helpful discussions with H. M. Weng, Y. G.
Yao, C. C. Liu and J. J. Zhou. This work was supported by the NKBRSFC under
grants Nos. 2011CB921502, 2012CB821305, 2009CB930701, 2010CB922904, NSFC
under grants Nos. 10934010, 11228409, 61227902 and NSFC-RGC under grants
Nos. 11061160490 and 1386-N-HKU748/10.


\begin{thebibliography}{99}
\bibitem{deltaE} $\Delta E=E_{s}+E_{ads}-E_{s-ads}$ is the adsorption energy
within GGA, where $E_{s}$, $E_{ads}$, $E_{s-ads}$ are energies of the $%
4\times 4$ pristine silicene, single adatom, and silicene-adatom system,
respectively.
\end{thebibliography}
\end{document}

\title{Supplemental materials for "Quantum anomalous Hall effect and tunable
topological states in $3d$ transition metals doped silicene"}
\author{Xiao-Long Zhang}
\affiliation{Beijing National Laboratory for Condensed Matter Physics,
Institute of Physics, Chinese Academy of Sciences, Beijing 100190, China}
\author{Lan-Feng Liu}
\affiliation{Beijing National Laboratory for Condensed Matter Physics,
Institute of Physics, Chinese Academy of Sciences, Beijing 100190, China}
\author{Wu-Ming Liu}
\affiliation{Beijing National Laboratory for Condensed Matter Physics,
Institute of Physics, Chinese Academy of Sciences, Beijing 100190, China}
\date{\today }
\maketitle

\section{Computational method}

The first-principles calculations are performed based on the density
functional theory (DFT) \cite{DFT} with generalized gradient approximation
(GGA) in the form of Perdew-Burke-Ernzerhof (PBE) functional \cite{PBE} as
implemented in Vienna Ab-initio Simulation Package (VASP) \cite{vasp}. The
GGA $+U$ method which treats the on-site repulsion interactions of $3d$
electrons in a mean field manner is used to evaluate the strong correlation
effect in TM, and a typical value of $U=4$ eV and $J=0.9$ eV are used for
all TM concerned  \cite{Uvalue}. The lattice constant $a$ $=3.86$
$\mathring{A}$ of silicene and the buckling distance $\delta =0.44$ $\mathring{A}$ are
obtained corresponding to the global minima on the Born-Oppenheimer surface,
which agree with existing theoretical data \cite{silicene, LCCL}. As to the
structure relaxation, all atoms are allowed to relax freely along all
directions and all parameters are chosen to converge the forces to less than
$0.01$ eV/$\mathring{A}$. A vacuum space of $20$ $\mathring{A}$ is set to
prevent the interaction between silicene and its periodic images along $c$%
-axis. Convergence tests with respect to energy cutoff and k points sampling
are performed to ensure numerical accuracy of total energy. We find an
energy cutoff of $420$ eV and $\Gamma $ centered Monkhorst-Pack grids of $%
8\times 8\times 1$ for k point sampling are enough to converge the
difference in total energy to around $1m$eV.

\section{impact of strong correlation effect on Adsorption structure and magnetic moments}

For the sake of comparison, we define bond lengths ($d_{Si-TM}$)
as the distance between adatom and the 3 nearest neighbour Si atoms to it, adsorption height ($h_{Si-TM}$) as the distance between adatom and the
lower sublattice (averaged) in $c$-axis for all 3 adsorption sites, height
above Si atom ($\delta _{Si-TM}$) as the distance between adatom and the Si
atom underneath\ for T$_{A}$ and T$_{B}$ sites (Fig. 1).
\begin{figure}[tbph]
\includegraphics[width = 9.0cm]{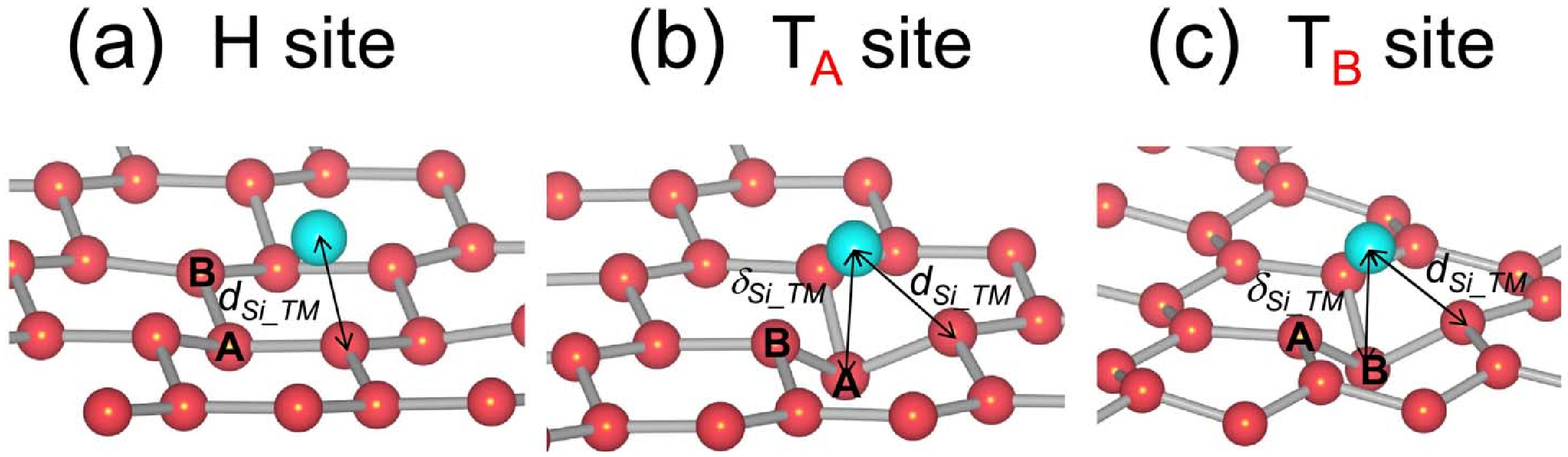}
\caption{{}Schematic representation of bond parameters ($d_{Si-TM}$ and $%
\protect\delta _{Si-TM}$) for 3 adsorption sites (a) H, (b) T$_{A}$,
and (c) T$_{B}$, respectively. }
\end{figure}

\begin{figure}[tbph]
\includegraphics[width = 8cm]{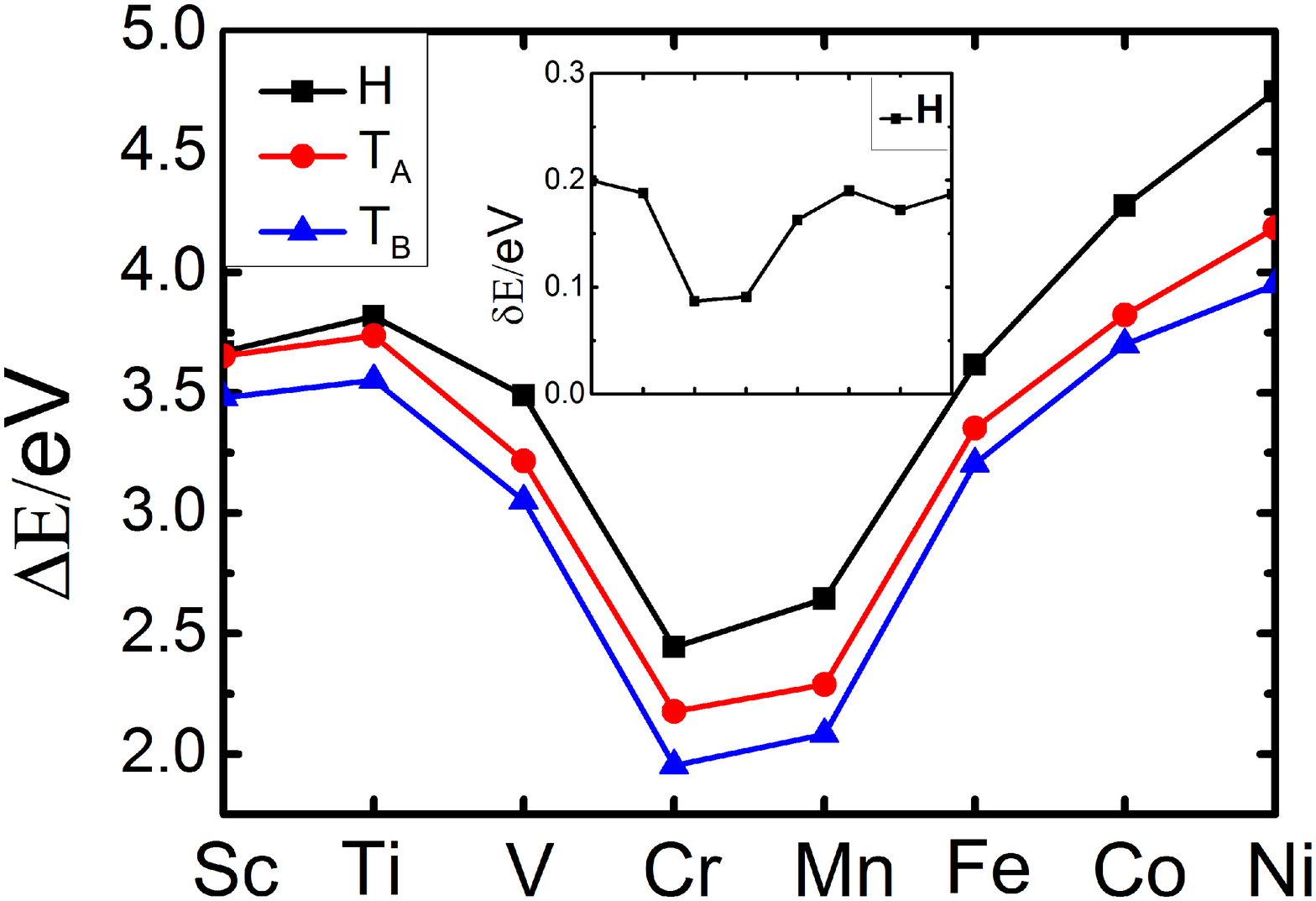}
\caption{{}The adsorption energy ($\Delta E$) of all $3d$ transition metals
(TM) adsorbed on the three high symmetric sites of silicene monolayer, H, T$%
_{A}$ and T$_{B},$ in the GGA level. The inset shows the distortion energy ($%
\protect\delta E$) of silicene when different TM are situated on the stable
site H.}
\end{figure}

\begin{figure}[tbph]
\includegraphics[width = 8.8cm,height = 9.5cm]{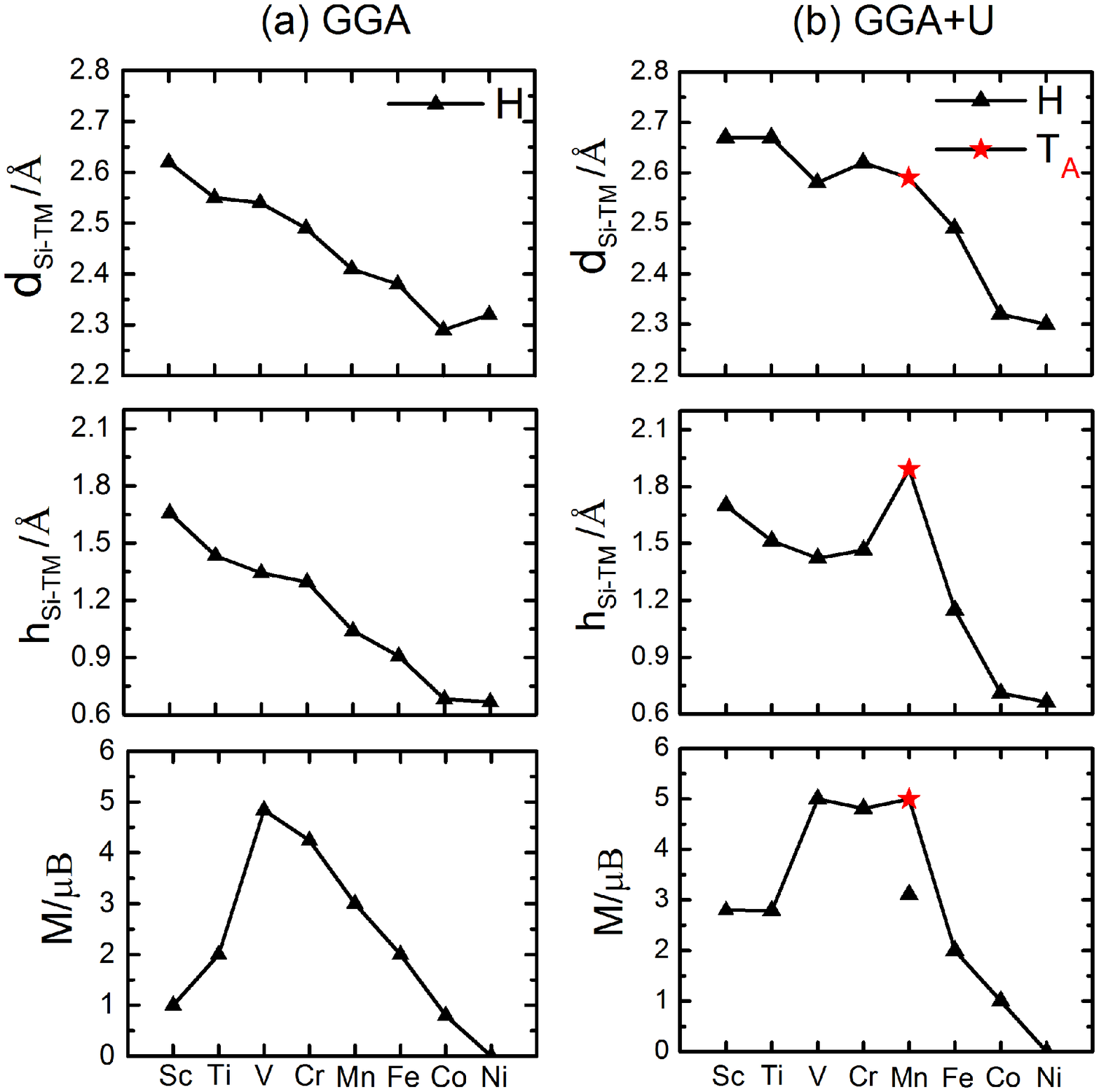}
\caption{{}Structural and magnetic properties of $3d$ transition metals
adsorbed on the Hollow site of $4\times 4$ silicene based on (a) GGA and (b)
GGA$+U$. For Mn in GGA$+U$ case, the bond parameters are corresponding to the stablest T$_{A}$
adsorption site (marked in red pentacle), and the magnetic moment of Hollow and T$_{A}$ sites are given for comparison.}
\end{figure}

In the GGA case, as can be seen from Fig. 3(a), the bond lengths and adsorption height
generally decrease with increasing of atomic number when TM adsorbing on H
site. However, the adsorption energy doesn't follow this trend (Fig. 2),
which has minimal value of $2.44$ eV for Cr and maximal value of $4.75$ eV
for Ni. We can see from inset in Fig. 2 that the above trend of adsorption
energy is correlated to the different distortion of silicene, which is
energetically characterized by the distortion energy defined as $\delta
E=E_{dis}-$ $E_{s}$, where $E_{dis}$ is the energy of silicene after
adsorption. The distortion energy are small for V and Cr, suggesting
relatively weak interactions between these adatoms and silicene, therefore,
the adsorption energy decrease from V and reaches minimal value at Cr.

As can be seen from Fig. 5, when Sc, Ti, Cr adsorbing on the H site, the
density of states (DOS) show peaks at the Fermi level, suggesting
a possible Jahn-Teller
distortion. In the case of
Sc-silicene, we artificially move one of three Si atoms nearest to Sc to break the $C_{3}$ rotational symmetry. After relaxation the 3
nearest Si atoms to Sc which originally coplanarly arranged themselves as a
regular triangle ($d_{Si-Sc}$ equal to $2.62\mathring{A}$.) now distort to a
isosceles triangle ($d_{Si-Sc}\ $become $2.63$ $\mathring{A},2.63$ $%
\mathring{A}$ and $3.12$ $\mathring{A}$.) by pushing the moved Si
atom down away from the upper sublattice plane by $1.24$ $\mathring{A}$.\
The distorted Sc-silicene system becomes more stable than $C_{3v}$ symmetric
one by lowering the total energy by 0.1eV. Similar to Sc-silicene case, we
could expect Jahn-teller distortion to further stabilise Ti-silicene and
Cr-silicene systems, nevertheless, the distortion for these two system
turned out to be rather weak (The modification of position of all atoms is
less than $0.005$ $\mathring{A}$, and total energy of distorted system which
no longer respect the $C_{3v}$ symmetry is lower by $\sim 3m$eV).

The resulting magnetic
moments and magnetic instability aforementioned can be attributed to the interplay
among crystal field splitting, the spin splitting, as well as the electron
occupation number of isolated $3d$ adatoms. For Sc-silicene, the spin splitting for Sc is relatively weak (around 0.2 eV), which is smaller than ligand field splitting between $E_{2}$ and $A_{1}$, and
totally there are 3 electrons occupying 3d orbitals (see Fig. 5). Therefore, two of these
3 electrons occupy the majority $E_{2}$ orbitals and the other one occupies
doubly degenerate minority $E_{2}$ orbitals, leading to 1 $\mu _{B}$
magnetic moment and potential Jahn-Teller distortion discussed above. For Ti-silicene, owing to the
relatively large splitting of $A_{1}$ ($1$eV), the majority $%
A_{1}$ orbital is occupied before the doubly degenerate minority $E_{2}$
orbitals as indicated in inset of Ti's projected density of states (PDOS) from GGA (Fig.
5), resulting in peaks at Fermi level and magnetic moment of 2 $\mu _{B}$.
For V, the spin splitting is much larger and high spin state with $5$ $\mu
_{B}$ moment state is realized, which is essential to the realization of
QAHE in silicene as has been discussed before. The
other cases can be understood in similar arguments.

When turning on the strong correlation effect, the equilibrium structure of
adatom-silicene systems are strongly altered compared with GGA case. As can
be seen from Fig. 3(b), the $d_{Si-TM}$, $h_{Si-TM}$ and $\delta _{Si-TM}$
for all adsorbates (except Ni) are increased, especially for Ti, Cr, Mn, Fe
(the bond lengths for these atoms increased by $\sim 0.1$ $\mathring{A}$
while for others by $\sim 0.05$ $\mathring{A}$, and the adsorption height
also showed noticeable rise for these atoms). And the H site is still
favored by most $3d$ TM (except Mn, which energetically favors T$_{A}$ site).

The change of adsorption geometry of adatom-silicene system can be
attributed to the direct consequence of on-site Coulomb interactions among $%
3d$ electrons. In the case of Sc-silicene, the GGA + $U$ predicts $\sim 3$ $\mu _{B}$ magnetic moment
compared with $\sim 1$ $\mu _{B}$ in GGA case (Fig. 3). And the reason for this is the enhanced spin splitting, which results from the effective $U$, makes
spin-up $A_{1}$ and $E_{2}$ states of Sc occupied as indicated in PDOS of Sc in Fig. 6.
For Ti-silicene system, the net magnetic
moment is $\sim 3$ $\mu _{B}$\ rather than $\sim 4$ $\mu _{B}$\ as one may
expected, indicating the $4s$\ shell of Ti is empty and about one electron
is transferred from Ti to silicene. This is justified by the lowering of
Dirac cone at $K$ and $-K$, though slightly distorted, beneath the Fermi
level, and occupied PDOS of Si atoms closest
to Ti. Similar to Ti-silicene, the Cr-silicene system retained $\sim 5$ $\mu
_{B}$ magnetic moment after transferring $\sim 1e$ to silicene, which shifts
the Dirac cone down by $\sim 0.4$ eV relative to Fermi level. In the case of
Mn-silicene, especially, the T$_{A}$ site is favored if the strong
correlation effect is turned on, which is 0.07 eV (0.13 eV) lower in total
energy than H (T$_{A}$) sites. For V, Fe, Co, Ni, $+U$ mainly enhances the
spin splitting while leaving the electron distribution nearly unaffected
compared with GGA case, hence the same magnetic moments (Fig. 3).

\begin{figure}[tbph]
\includegraphics[width = 9.0cm]{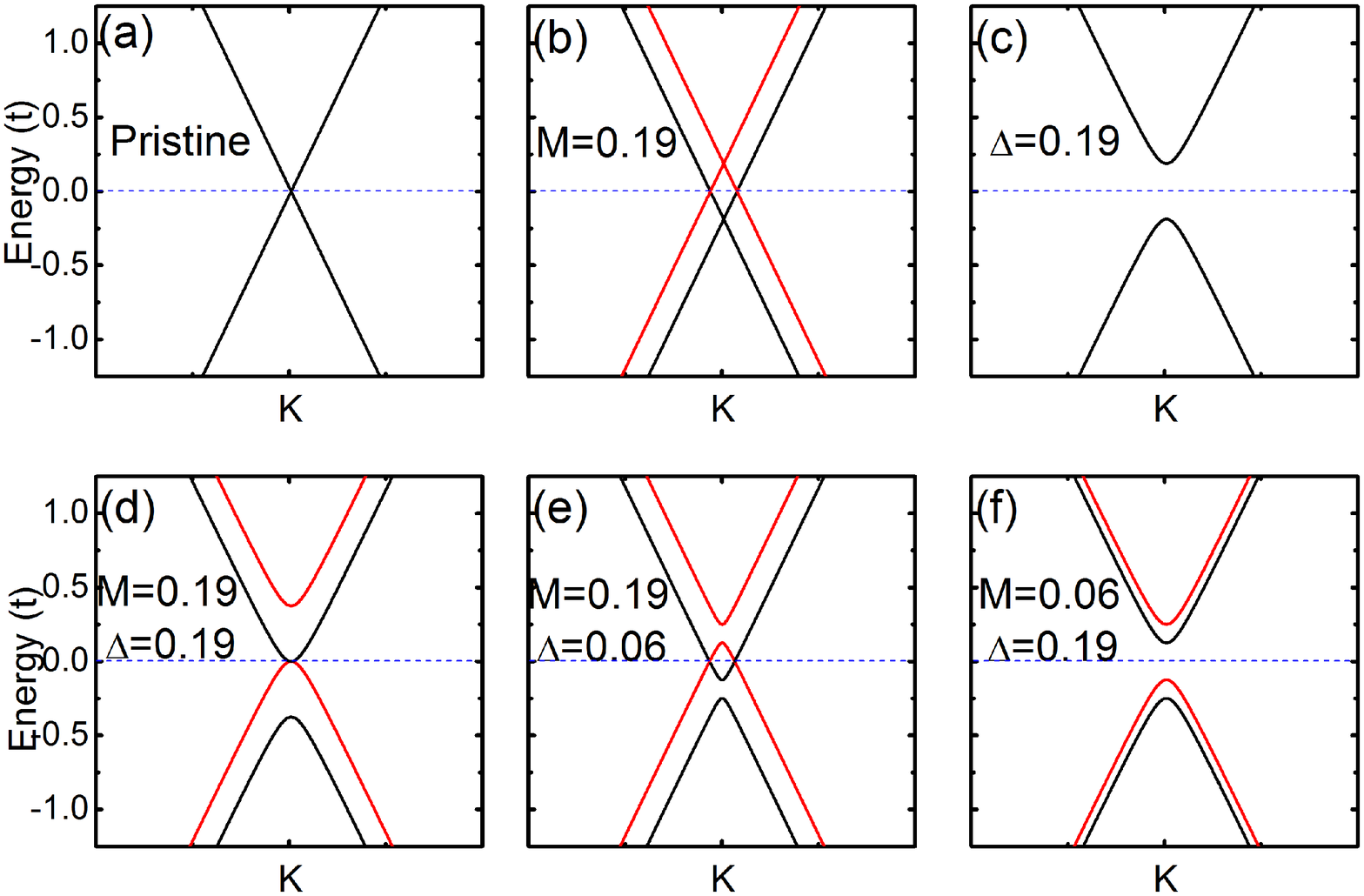}
\caption{{}The evolution of band structure around valley $K$ from the interplay between
exchange field $M$ and staggered potential $\Delta $ (in unit of $t$). The
red (black) lines are for the majority (minority) spin.}
\end{figure}
\begin{figure*}[tbph]
\includegraphics[width = 18.0cm]{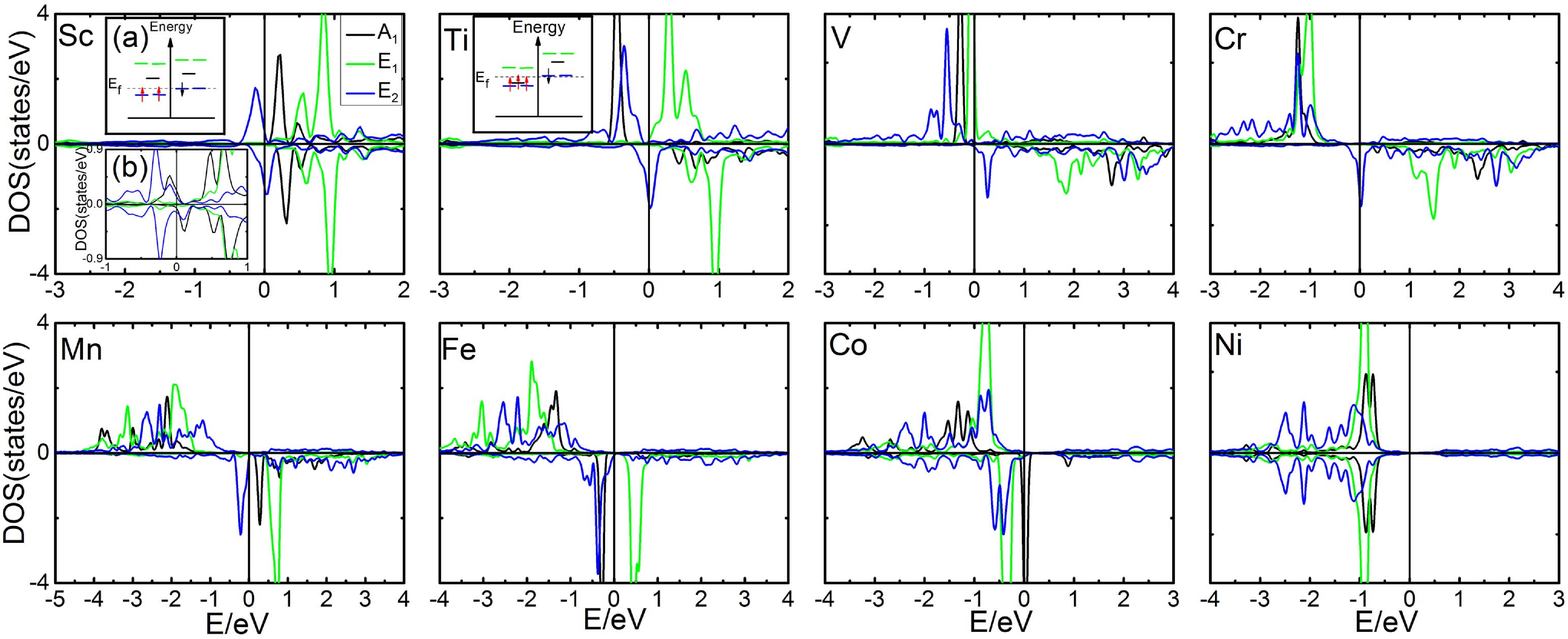}
\caption{{}PDOS of all $3d$ transition metals
adsorbed on the stable site (Hollow) of monolayer silicene from GGA, where
positive (negative) values are for majority (minority) spin. The inset (b) in Sc indicates the Jahn-Teller distorted PDOS. The Fermi energy is set to 0 eV.}
\end{figure*}

\begin{figure*}[tbph]
\includegraphics[width =18.0cm]{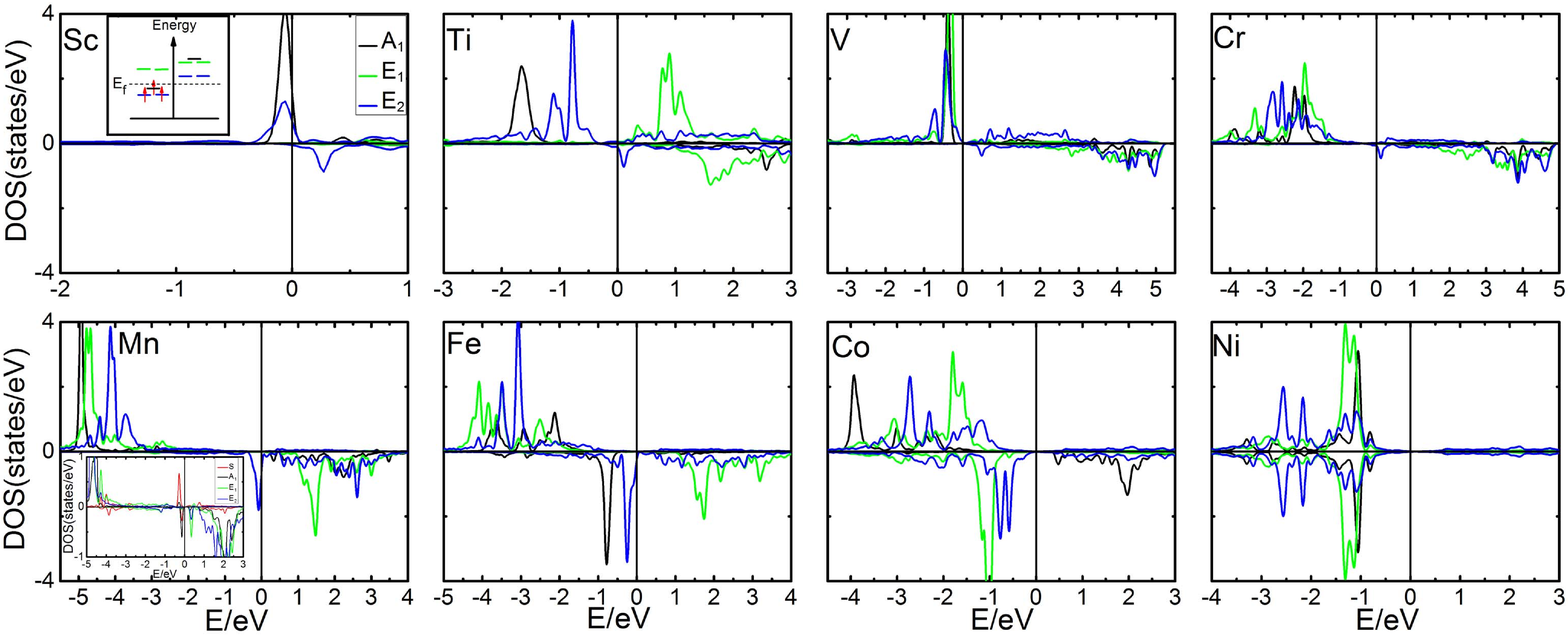}
\caption{PDOS of all $3d$ transition metals
adsorbed on the Hollow (H) site of monolayer silicene from GGA$+U$ $($with
an effective $U$ being $3.1eV$), where positive (negative) values are for
majority (minority) spin. For Mn, the PDOS including $4s$ orbitals are given
in inset when absorbing on the stable site T$_{A}.$ The Fermi energy is set
to 0 eV.}
\end{figure*}
\section{Band structure evolution from Tight binding model}

Fig. $4$ illustrates the evolution of the band structure around valley $K$
with different combination of exchange field $M$ and staggered potential $\Delta$. Panel(a) shows the band structure
of pristine silicene with perfect Dirac-like energy dispersion. In panel(b),
the spin degeneracy is lifted when only exchange field $M$ is turned on,
rendering crossing of bands with opposite spin around $K$ point in the
reciprocal space similar to graphene case \cite{Qiao rapid}. While if only
staggered potential $\Delta $\ is applied (panel(c)), the system becomes
insulating with the valence and conduction bands twofold degenerated.

As mentioned before, in $3d$ TM doped silicene, the exchange field and
staggered potential should appear simultaneously. Panels(d)-(e) indicate the
interplay between exchange field ($M$) and staggered potential ($\Delta $).
In the case of panel(d) where $M=\Delta $, there always exists a degenerate point
right at the Fermi level. When increasing $M$ and making $M>\Delta$ (panel(e)), the
two spin subbands near Fermi level cross, resulting a circular Fermi
surface. While if keeps increase exchange field $M$ (panel(f)), the system enters
insulating state. The $M>\Delta $ case is of particular interest in our
investigation, since upon turning on the spin orbit coupling effect, the case (e)
will give rise to the QAHE and electrically tunable topological states from QAHE to QVHE as discussed before.